\documentclass[twoside,11pt]{article}

\usepackage[accepted, abbrvbib]{melba}

\usepackage{amsmath,amsfonts}

\usepackage{mathtools}
\usepackage[ruled]{algorithm2e}
\usepackage{algpseudocode}

\usepackage{multirow}
\usepackage{subfig}

\melbaheading{2022:004}{https://www.melba-journal.org/papers/2022:004.html}{2022}{1-29}{09/2021}{02/2022}{Saeed, Yan, Fu, Giganti, Yang, Baum, Rusu, Fan, Sonn, Emberton, Barratt and Hu}{IPMI 2021}{Aasa Feragen, Stefan Sommer, Mads Nielsen and Julia Schnabel}

\ShortHeadings{Image quality assessment by overlapping task-specific and task-agnostic measures}{Saeed, Yan, Fu, Giganti, Yang, Baum, Rusu, Fan, Sonn, Emberton, Barratt, Hu}
\firstpageno{1}

\title{Image quality assessment by overlapping task-specific and task-agnostic measures: application to prostate multiparametric MR images for cancer segmentation}

\author{\name Shaheer U. Saeed 
    \email shaheer.saeed.17@ucl.ac.uk \\  
    \addr Centre for Medical Image Computing; Wellcome/EPSRC Centre for Interventional \& Surgical Sciences; Department of Medical Physics \& Biomedical Engineering, University College London, London, UK.
	\AND
	\name Wen Yan\\
	\addr Centre for Medical Image Computing; Wellcome/EPSRC Centre for Interventional \& Surgical Sciences; and Department of Medical Physics \& Biomedical Engineering, University College London, London, UK; City University of Hong Kong, Department of Electrical Engineering, Hong Kong, China.
	\AND
	\name Yunguan Fu\\
	\addr Centre for Medical Image Computing; Wellcome/EPSRC Centre for Interventional \& Surgical Sciences; and Department of Medical Physics \& Biomedical Engineering, University College London; and InstaDeep, London, UK.
	\AND
	\name Francesco Giganti\\
	\addr Department of Radiology, University College London Hospital NHS Foundation Trust; and Division of Surgery and Interventional Science, University College London, London, UK.
	\AND
	\name Qianye Yang\\
	\addr Centre for Medical Image Computing; Wellcome/EPSRC Centre for Interventional \& Surgical Sciences; Department of Medical Physics \& Biomedical Engineering, University College London, London, UK.
	\AND
	\name Zachary M. C. Baum\\
	\addr Centre for Medical Image Computing; Wellcome/EPSRC Centre for Interventional \& Surgical Sciences; Department of Medical Physics \& Biomedical Engineering, University College London, London, UK.
	\AND
	\name Mirabela Rusu\\
	\addr Department of Radiology, Stanford University, Stanford, California, USA.
	\AND
	\name Richard E. Fan\\
	\addr Department of Urology, Stanford University, Stanford, California, USA.
	\AND
	\name Geoffrey A. Sonn\\
	\addr Department of Radiology; and Department of Urology, Stanford University, Stanford, California, USA.
	\AND
	\name Mark Emberton\\
	\addr Department of Urology, University College  Hospital NHS Foundation Trust; and Division of Surgery and Interventional Science, University College London, London, UK.
	\AND
	\name Dean C. Barratt\\
	\addr Centre for Medical Image Computing; Wellcome/EPSRC Centre for Interventional \& Surgical Sciences; Department of Medical Physics \& Biomedical Engineering, University College London, London, UK.
	\AND
	\name Yipeng Hu\\
	\addr Centre for Medical Image Computing; Wellcome/EPSRC Centre for Interventional \& Surgical Sciences; Department of Medical Physics \& Biomedical Engineering, University College London, London, UK.
}

\begin{document}

\maketitle

\begin{abstract}
Image quality assessment (IQA) in medical imaging can be used to ensure that downstream clinical tasks can be reliably performed. Quantifying the impact of an image on the specific target tasks, also named as task amenability, is needed. 
A \textit{task-specific} IQA has recently been proposed to learn an image-amenability-predicting controller simultaneously with a target task predictor. This allows for the trained IQA controller to measure the impact an image has on the target task performance, when this task is performed using the predictor, e.g. segmentation and classification neural networks in modern clinical applications. In this work, we propose an extension to this task-specific IQA approach, by adding a \textit{task-agnostic} IQA based on auto-encoding as the target task~\footnote{Although the auto-encoding for image self-reconstruction can be considered a `target task' in the training algorithm (Section~\ref{sec:task_agnos_iqa}), it is not a clinical task. This target task provides clinical-task-agnostic reward signals for training the task-agnostic IQA.}. Analysing the intersection between low-quality images, deemed by both the task-specific and task-agnostic IQA, may help to differentiate the underpinning factors that caused the poor target task performance. For example, common imaging artefacts may not adversely affect the target task, which would lead to a low task-agnostic quality and a high task-specific quality, whilst individual cases considered clinically challenging, which can not be improved by better imaging equipment or protocols, is likely to result in a high task-agnostic quality but a low task-specific quality.
We first describe a flexible reward shaping strategy which allows for the adjustment of weighting between task-agnostic and task-specific quality scoring. Furthermore, we evaluate the proposed reinforcement learning algorithm, using a clinically challenging target task of prostate tumour segmentation on multiparametric magnetic resonance (mpMR) images. Based on experimental results using mpMR images from 850 patients, it was found that \textit{a)} The task-agnostic IQA may identify artefacts, but with limited impact on the accuracy of cancer segmentation networks. A Dice score of $0.367\pm0.017$ was obtained after rejecting 10\% of low quality images, compared to $0.354\pm0.016$ from a non-selective baseline; \textit{b)} The task-specific IQA alone improved the performance to $0.415\pm0.020$, at the same rejection ratio. However, this system indeed rejected both images that impact task performance due to imaging defects and due to being clinically challenging; and \textit{c)} The proposed reward shaping strategy, when the task-agnostic and task-specific IQA are weighted appropriately, successfully identified samples that need re-acquisition due to defected imaging process, as opposed to clinically challenging cases due to low contrast in pathological tissues or other equivocacy in radiological presentation.

Our code is available at~\url{https://github.com/s-sd/task-amenability/tree/v1}.
\end{abstract}



\section{Introduction}\label{sec:intro}

\subsection{Image quality assessment}

Image quality assessment (IQA) is utilised extensively in medical imaging and can help to ensure that intended downstream clinical tasks for medical images, such as diagnostic, navigational or therapeutic tasks, can be reliably performed. It has been demonstrated that poor image quality may adversely impact task performance \citep{qa_retina, qa_fetal, qa_review}. In these scenarios IQA can help to ensure that the negatively impacted performance can be counter-acted for example by flagging samples for re-acquisition or defect correction. Most medical images are acquired with an intended clinical task, however, despite the task-dependent nature of medical images, IQA is often studied in task-agnostic settings \citep{qa_review}, in which the task is independently performed or, for automated tasks, the task predictors (e.g. machine learning models) are disregarded. Where task-specific IQA, which accounts for the impact of an image on a downstream task, is studied, the impact is usually quantified subjectively by human observers or is learnt from subjective human labels.

Both manual and automated methods have been proposed for task-agnostic as well as task-specific IQA. In clinical practice, manually assessing images for their perceived impact on the target task is common practice \citep{qa_review}. Usually, manual assessment involves subjectively defined criteria used to assess images where the criteria may or may not account for the impact of image quality on the target task \citep{qa_review, qa_subj_single_2, qa_subj_single_3, qa_subj_single_4, qa_subj_multiple_1, qa_subj_multiple_2}. Automated methods to IQA, either task-specific or task-agnostic, enable reproducible measurements and reduce subjective human interpretation \citep{qa_review}. So called no-reference automated methods capture image features common across low or high quality images, in order to perform IQA \citep{qa_no_ref_1, qa_no_ref_3, qa_no_ref_4, qa_no_ref_7}. The selected common features may be task-specific in some cases, however, these methods are often task-agnostic since the kinds of features selected in general have no bearing on a specific clinical task (e.g. distortion lying outside gland boundaries for gland segmentation). Full- and reduced-reference automated methods use a single or a set of reference images and produce IQA measurements by quantifying similarity to the reference images \citep{qa_ref_sim_1, qa_ref_sim_10, qa_ref_sim_11, qa_ref_sim_12, qa_ref_sim_13, qa_ref_sim_14, qa_ref_sim_15, qa_ref_sim_16, qa_ref_sim_17, qa_ref_sim_2, qa_ref_sim_3, qa_ref_sim_4, qa_ref_sim_5, qa_ref_sim_6, qa_ref_sim_7, qa_ref_sim_5, qa_ref_sim_8, qa_ref_sim_9}. The selection of the reference, to which similarity is computed, is usually either done manually or using a set of subjectively defined criteria which may or may not account for impact on a clinical task.

Recent deep learning based approaches to IQA also rely on human labels of IQA and offer fast inference \citep{qa_fetal, qa_retina_deep, qa_liver_deep, qa_zachary_lung, liao_qa_ec, abdi_qa_ec, lin_qa_multitask, oksuz_mr_iqa}. Although these methods can provide reproducible and repeatable measurements, the extent to which they can quantify the impact of an image on the target task remains unanswered. 

Perhaps more importantly, many downstream clinical target tasks have been automated using machine learning models, and therefore the human-perceived quality, even when they are intended to be task-specific \citep{qa_fetal, qa_liver_deep, qa_no_ref_task_specific_1, qa_no_ref_task_specific_2}, may not be representative of the actual impact of an image on the task to be completed by computational models. This is also true for subjective judgement that is involved in other manual or automated no-, reduced- or full-reference methods. Some works do objectively quantify image impact on a downstream task, however, they are specific to particular applications or modalities,  for example for IQA or under-sampling in MR images \citep{qa_no_ref_4, qa_no_ref_7, razumov_undersampling_mr}, IQA for CT images \citep{qa_no_ref_task_specific_1, qa_no_ref_task_specific_2} and synthetic data selection for data augmentation \citep{ye_synthetic_rl}. In a typical scenario where imaging artefacts lie outside regions of interest for a specific task, general quantification of IQA may not indicate usefulness of the images. For a target task of tumour segmentation on MR images, if modality-specific artefacts, such as motion artefact and magnetic field distortion, are present but of great distance from the gland, then segmentation performance may not be impacted negatively by these artefacts. We present examples of this type in Sect. 4.

\subsection{Task-specific image quality: task amenability}

In our previous work \citep{saeed_amenability} we proposed to use the term `task amenability' to define the usefulness of an image for a particular target task. This is a task-specific measure of IQA, by which selection of images based on their task amenability may lead to improved target task performance. Task amenability-based selection of images may be useful under several potential scenarios such as for meeting a clinically defined requirement on task performance by removing images with poor task amenability, filtering images in order to re-acquire for cases where the target task cannot be reliably performed due to poor quality, and real-time acquisition quality or operator skill feedback. `Task amenability' is used interchangeably with `task-specific image quality' in this paper henceforth.

In order to quantify task amenability, a controller, which selects task amenable image data, and a task predictor, which performs the target task, are jointly trained. In this formulation the controller decisions are informed directly by the task performance, such that optimising the controller is dependent on the task predictor being optimised - a meta-learning problem. We thus formulated a reinforcement learning (RL) framework for this meta-learning problem, where data weighting/selection is performed in order to optimise target task performance \citep{saeed_amenability}. 

Modifying meta- or hyper-parameters to optimise task performance using RL algorithms has been proposed previously, including selecting augmentation policies \citep{autoaugment, adv_autoaugment}, selecting convolutional filters and other neural network hyper-parameters \citep{archisearch}, weighting different imaging modalities \citep{autoweight}, and valuating or weighting training data \citep{google_dvrl}; the target task may be any task performed by a neural network, for example classification or regression. The independently-proposed task amenability method \citep{saeed_amenability} shares some similarity with the data valuation approach \citep{google_dvrl}, however, different from \citet{google_dvrl}, our work investigated reward formulations using controller weighted/ selected data and the use of the controller on holdout data, in addition to other methodological differences.

\subsection{Intersection between task-specific and task-agnostic low-quality images}\label{sec:overlap}

\begin{figure}
    \centering
    \includegraphics[width=\textwidth]{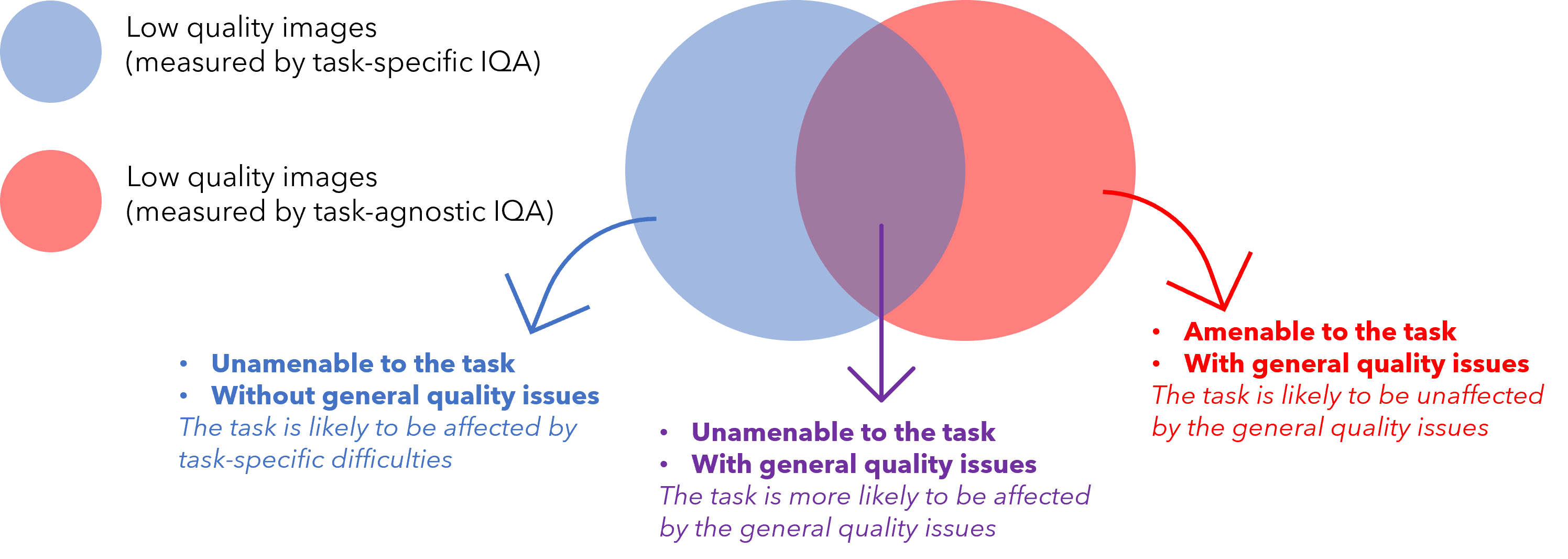}
    \caption{Venn diagram to illustrate the low quality images with respect to task-specific and task-agnostic IQA measures, and their intersections. }
    \label{fig:venn_iqa}
\end{figure}

It is worth noting that our previously proposed framework is task-specific and therefore may not be able to distinguish between poor-quality samples due to imaging defects and those due to clinical difficulty. For example, for the diagnostic task of tumour segmentation, despite no visible imaging artefacts, an image may still be considered challenging because of low tissue contrast between the tumour and surrounding regions or because of a very small tumour size. It may have a different outcome compared to cases where imaging defects, such as artefacts or noise, cause reduced task performance. The former cases that are clinically challenging may require further expert input or consensus, whereas latter cases with imaging defects may need to be re-acquired. 

Therefore, it may be useful to perform task-agnostic IQA, when imaging defects need to be identified. Task-agnostic methods to IQA quantify image quality as a metric based on the presence of severe artefacts, excessive noise or other general imaging traits for low quality. Those observable general quality issues include unexpected low spatial resolution, electromagnetic noise patterns, motion-caused blurring and distortion due to imperfect reconstruction, which often determine human labels for poor task-agnostic quality. Although these may not affect the target task and/or without necessarily well-understood underlying causes or definitive explanations, these task-agnostic quality issues should be flagged to indicate potentially avoidable or amendable imaging defects. Examples of previous work include those using auto-encoders as feature extractors for clustering or using supervised learning \citep{autoenc_iqa_cluster, autoenc_iqa_supervised}. 

Fig \ref{fig:venn_iqa} illustrates the relationship between task-specific and task-agnostic qualities discussed in this work as a Venn diagram. Task-specific IQA identifies all samples (blue circle) that negatively impact task performance regardless of the underlying reason. In contrast, task-agnostic IQA may be able to identify images (red circle) with general quality issues, as discussed above, regardless of their impact on the target task. Therefore, image samples that are identified by the task-specific IQA but \textit{not} by the task-agnostic IQA (blue area excluding the purple overlap) are likely to be without readily-identifiable general imaging quality issues, thus more likely to be difficult cases due to patient-specific anatomy and pathology. Whilst, image samples that suffer from general quality issues, such as imaging defects, \textit{and} also negatively impact task performance shall lie in the (purple) overlap between samples identified by both task-specific- and task-agnostic IQA.

In this work, we propose to incorporate task-agnostic IQA by a flexible reward shaping strategy, which allows for the weighting between the two IQA types to be adjusted for the application of interest. To valuate images based on their task-agnostic image quality, we propose to use auto-encoding as the target task, trained in the same RL algorithm. It is trained with a reward based on image reconstruction error, since random noise and rare artefacts may not be reconstructed well by auto-encoders \citep{autoenc_denoise_1, autoenc_denoise_2, autoenc_denoise_med}. Once such a task-agnostic IQA is learned, it can be used to shape the originally-proposed task-specific reward. This allows to train a final IQA controller that ranks the images by both task-specific and task-agnostic measures, with which the lowest-ranked images are much more likely to be those depicted in the overlap area of Fig \ref{fig:venn_iqa}, compared to either of these two IQA approaches alone. A completely task-specific IQA controller and a completely task-agnostic IQA controller are both a special case of the proposed reward shaping mechanism, by configuring a hyper-parameter that controls this quality-ranking.

\subsection{Multiparametric MR quality for prostate cancer segmentation}

Prostate cancer is among the most commonly occurring malignancies in the world \citep{prostate_cancer}. In current prostate cancer patient care, biopsy-based histopathology outcome remains the diagnostic gold-standard. However, the emergence of multiparametric magnetic resonance (mpMR) imaging does not only offer a potential non-invasive alternative, but also a role in better localising tumour for targeted biopsy or other increasingly localised treatment, such as several focal ablation options \citep{prostate_manual_tumour_seg, ahmed_focal_1, marshall_focal_2, ahdoot_focal_3}.

Detecting, grading and segmenting tumours from prostate mpMR are known challenging radiological tasks, with a reported 7-14\% of missed clinically significant cancers \citep{ahmed_promis, rouviere_mri_first}, a high inter-observer variability \citep{prostate_manual_tumour_seg} and a strong dependency on the image quality \citep{de_mpMR_iqa}, echoed by ongoing effort in standardising both the radiologist reporting \citep{weinreb_pi_rads} and the IQA for mpMR \citep{giganti_pi_qual}. 

The clinical challenges may directly lead to high-variance in both radiological and histopathological labels, for training machine learning models to aid this diagnostic task. Recently proposed methods \citep{prostate_auto_tumour_seg, prostate_ml_tumour_seg, prostate_ml_tumour_seg_2}, mostly based on deep neural networks, reported a relatively low segmentation accuracy in Dice ranging from 20\% to 40\%. When analysing these machine learning models to improve the target task performance, the two types of image quality issues discussed in Sect.~\ref{sec:overlap} have both been observed. Common image quality issues such as grain, susceptibility or distortion \citep{iqa_prostate_mr} can adversely impact lesion segmentation, so can a small or ambiguous lesion at early stage of its malignancy. The desirable image quality is also likely to be different for different target tasks, for example gland segmentation \citep{soerensen_gland_seg, ghavami_gland_seg}. As discussed in previous sections, a re-scan at an experienced cancer centre with a well-tuned mpMR protocol may help certain type of patient cases, but others may require a second radiologist reading or further biopsy. 

In this work, we use this clinically-difficult and imaging-dependent target task to test the proposed IQA system, 1) for assessing the mpMR image quality and 2) for further sub-typing the predicted quality issues. As discussed above, both of these abilities are important in informing subsequent clinical decisions, when these machine learning models are deployed in prostate cancer patient care.

\subsection{Contribution summary}

This paper extends the previously published work~\citep{saeed_amenability} and includes original contributions as follows.
\textbf{First}, 1a) we review the proposed IQA formulation based on task amenability; and 1b) summarise the results based on a previously described interventional application, in which transrectal ultrasound (TRUS) image quality was evaluated with respect to two different target tasks, classifying and segmenting prostate glands; and 1c) we further consolidate the proposed IQA approach with results from new ablation studies to better evaluate different components in the framework; 
\textbf{Second}, 2a) we describe a new extension to use an independently-learned task-agnostic IQA to augment the original task-specific formulation; and 2b) propose a novel flexible reward shaping strategy to learn such a weighted IQA using a single RL-based framework; and 2c) we describe a new set of experimental results, based on a new clinical application, deep-learning-based cancer segmentation from prostate mpMR images, in order to evaluate both the original and the extended IQA approaches.

\section{Methods}

\begin{figure}
    \centering
    \includegraphics[width=\textwidth]{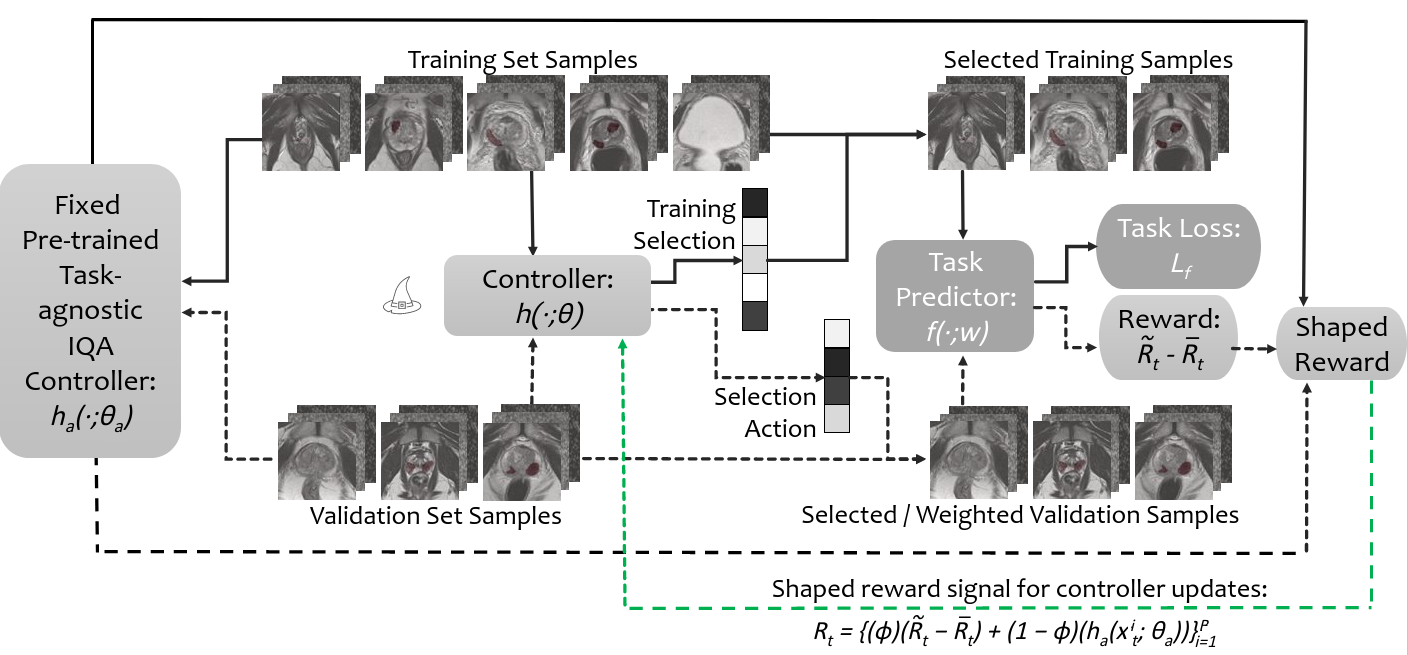}
    \caption{Summary of the IQA framework, with which all types of IQA controllers discussed in this work can be trained. The black solid lines indicate the pathway for the train set data and black dashed lines indicate the pathway for the validation set data; green dashed line indicates reward used to update controller). }
    \label{fig:methods_fig}
\end{figure}

\subsection{Learning image quality assessment}\label{sec:iqa}

\subsubsection{Image quality assessment}
The proposed image quality assessment framework is comprised of two parametric functions: 1) the task predictor, $f(\cdot;w):\mathcal{X}\rightarrow\mathcal{Y}$, which performs the target task by producing a prediction $y\in\mathcal{Y}$ for a given image sample $x\in\mathcal{X}$; 2) the controller, $h(\cdot;\theta):\mathcal{X}\rightarrow[0,1]$, which scores image samples $x$ based on their task-specific quality \footnote{Task-specific quality means quality scores that represent the impact of an image on a target task; the target task may include any machine learning task including classification, segmentation, or self-reconstruction where these task may not necessarily be clinical tasks.}. The image and label domains for the target task are denoted by $\mathcal{X}$ and $\mathcal{Y}$, respectively. Thus in this formulation, we define the image and joint image-label distributions as $\mathcal{P}_{X}$ and $\mathcal{P}_{XY}$, respectively. These distributions have probability density functions $p(x)$ and $p(x,y)$, respectively.

The loss function to be minimised for optimising the task predictor is $L_f:\mathcal{Y}\times\mathcal{Y}\rightarrow\mathbb{R}_{\geq0}$. This loss function $L_f$ measures how well the target task is performed by the task predictor $f(\cdot; w)$. This loss weighted by controller-predicted IQA scores can be minimised such that poor target task performance for image samples with low controller-predicted scores are weighted less:
\begin{align}
    \min_w \mathbb{E}_{(x,y)\sim\mathcal{P}_{XY}}[L_f(f(x;w), y)h(x;\theta)],
\end{align}

On the other hand the metric function $L_h:\mathcal{Y}\times\mathcal{Y}\rightarrow\mathbb{R}_{\geq0}$ measures task performance on a validation set. Intuitively, performing the task on images with lower task-specific quality tends to be difficult. To encourage the controller to predict lower quality scores for samples with higher metric values (lower task performance), the following weighted metric function can be minimised:
\begin{align}
    \min_\theta \mathbb{E}_{(x,y)\sim\mathcal{P}_{XY}}[L_h(f(x;w), y)h(x;\theta)],\\
    \text{s.t.}\quad \mathbb{E}_{x\sim\mathcal{P}_{X}}[h(x;\theta)] \geq c > 0
\end{align}

Here, $c$ is a constant that ensures non-zero quality scores.

The controller thus learns to predict task-specific quality scores for image samples while the task predictor learns to perform the target task. The IQA framework can thus be assembled as the following bi-level minimisation problem \citep{sinha_bilevel}:
\begin{subequations}\label{eq:iqa-weighted}
\begin{align}
    && \min_\theta \mathbb{E}_{(x,y)\sim\mathcal{P}_{XY}}[L_h(f(x;w^*), y)h(x;\theta)],\\
    \text{s.t.}&& w^*=\arg\min_w \mathbb{E}_{(x,y)\sim\mathcal{P}_{XY}}[L_f(f(x;w), y)h(x;\theta)],\\
    &&\mathbb{E}_{x\sim\mathcal{P}_{X}}[h(x;\theta)] \geq c > 0.
\end{align}
\end{subequations}

A re-formulation to permit sample selection based on task-specific quality scores, where the the data $x$ and $(x,y)$ are sampled from the controller-selected or -sampled distributions $\mathcal{P}_{X}^h$ and $\mathcal{P}_{XY}^h$, with probability density functions $p^h(x) \propto p(x)h(x;\theta)$ and $p^h(x,y) \propto p(x,y)h(x;\theta)$ respectively, can be defined as follows:
\begin{subequations}\label{eq:iqa-sampled}
\begin{align}
    && \min_\theta \mathbb{E}_{(x,y)\sim\mathcal{P}_{XY}^h}[L_h(f(x;w^*), y)],\\
    \text{s.t.}&& w^*=\arg\min_w \mathbb{E}_{(x,y)\sim\mathcal{P}_{XY}^h}[L_f(f(x;w), y)],\\
    &&\mathbb{E}_{x\sim\mathcal{P}_{X}^h}[1] \geq c > 0.
\end{align}
\end{subequations}

\subsubsection{Reinforcement learning}\label{sec:rl}

This bi-level minimisation problem can be formulated in a RL setting. We propose to formulate this problem in a RL-based meta learning setting where the controller interacting with the task predictor via sample-selection or -weighting can be considered a Markov decision process (MDP). The MDP in this problem setting can be described by a 5-tuple $(\mathcal{S}, \mathcal{A}, p, r, \pi)$. The state transition distribution $p:\mathcal{S}\times\mathcal{S}\times\mathcal{A}\rightarrow[0,1]$ denotes the probability of the next state $s_{t+1}$ given the current state $s_t$ and action $a_t$; here $s_{t+1},~s_t \in \mathcal{S}$ and $a_t \in \mathcal{A}$, where  $\mathcal{S}$ is the state space and $\mathcal{A}$ is the action space. The reward function is $r:\mathcal{S}\times\mathcal{A}\rightarrow\mathbb{R}$ and $R_t=r(s_t,a_t)$ denotes the reward given $s_t$ and $a_t$. The policy is $\pi(a_t\mid s_t):\mathcal{S}\times\mathcal{A}\in[0,1]$, which represents probability of performing action $a_t$ given $s_t$. The rewards accumulated starting from time-step $t$ can be denoted by: $Q^\pi(s_t, a_t) = \sum_{k=0}^{T}\gamma^kR_{t+k}$, where the discount factor for future rewards is $\gamma\in[0,1]$. A trajectory or sequence can be observed as the MDP interaction takes place, the sequence takes the form $(s_1, a_1, R_1, s_2, a_2, R_2, \ldots,  s_T, a_T, R_T)$. If the interactions take place according to a parameterised policy $\pi_\theta$, then the objective in RL is to learn optimal policy parameters $\theta^* = \text{argmax}_{\theta} \mathbb{E}_{\pi_\theta}[Q^\pi(s_t, a_t)]$.

\subsubsection{Predicting image quality using reinforcement learning}\label{sec:iqa_rl}

Formulating the IQA framework, outlined in Sect. \ref{sec:iqa}, as a RL-based meta-learning problem, the finite dataset together with the task predictor can be considered to be contained inside an environment with which the controller interacts. Then at time-step $t$, from the training dataset $\mathcal{D}_\text{train}=\{(x_t^i,y_t^i)\}_{i=1}^N$, a batch of samples $\mathcal{B}_t=\{(x_t^i,y_t^i)\}_{i=1}^B$ together with the task predictor $f(\cdot;w_t)$ may be considered an observed state $s_t=(f(\cdot;w_t), \mathcal{B}_t)$. The controller $h(\cdot;\theta)$ outputs sampling probabilities which dictate selection decisions for samples based on $a_t^i\sim\text{Bernoulli}(h(x_t^i; \theta))$ where action $a_t=\{a_t^i\}_{i=1}^B\in\{0,1\}^B$ contains the selection decision for samples in the batch. Since the actions are binary selection decisions, at each time-step $t$, the possible actions are $2^B$. If $a_t^i=1$, sample $(x_t^i,y_t^i)$ is selected for training the task predictor. The policy $\pi_\theta(a_t\mid s_t)$ may be defined as $\log\pi_\theta(a_t\mid s_t)=\sum_{i=1}^{D}h(x_t^i;\theta)a_t^i + (1-h(x_t^i;\theta)(1-a_t^i))$.

In this RL-based meta-learning framework, the reward is computed based on the task predictor's performance $\{l_{t}^j\}_{j=1}^M = \{L_h(f(x^j;w_t), y^j)\}_{j=1}^M$ on a validation set $\mathcal{D}_\text{val}=\{(x^j,y^j)\}_{j=1}^M$. The controller outputs for the validation set $\{h^j\}_{j=1}^M = \{h(x^j;\theta)\}_{j=1}^M$ may be utilised to formulate the final reward. While several strategies exist for reward computation, we investigate three strategies to compute an un-clipped reward $\tilde{R}_t$ Three strategies to compute $\tilde{R}_t$ are as follows:
\begin{enumerate}
    \item $\tilde{R}_{\text{avg},t}=-\frac{1}{M}\sum_{j=1}^Ml_{t}^j$, the average performance,
    \item $\tilde{R}_{\text{w},t}=-\frac{1}{M}\sum_{j=1}^Ml_{t}^j h^j$, the weighted sum,
    \item $\tilde{R}_{\text{sel},t}=-\frac{1}{M'}\sum_{j'=1}^{M'}l_{t}^{j'}$, the average of the selected $M'$ samples;
\end{enumerate}
For the selective reward formulation $\tilde{R}_{\text{sel},t}$ $\{{j'}\}_{j'=1}^{M'} \subseteq \{j\}_{j=1}^M$ with $h^{j'}\leq h^{k'}$ for $ \forall{k'\in\{j'\}^c}$, $\forall{j'\in\{j'\}}$, i.e. the unclipped reward $\tilde{R}_{\text{sel},t}$ is the average of $\{l_{j'}\}$ from the subset of $M'=\lfloor (1-s^{rej})M \rfloor$ samples, by removing the first $s^{rej}\times100\%$ samples, after sorting $h_j$ in decreasing order. The first reward formulation $\tilde{R}_{\text{avg},t}$ requires pre-selection of data with high task-specific quality whereas the other two reward formulations do not require any pre-selection of data and the validation set may be of mixed quality. In the selective and weighted reward formulations data is selected or weighted based on controller-predicted task-specific quality. The use of a validation set ensures that the system encourages generalisability therefore extreme cases, where the controller values one sample very highly and the remaining samples as very low, are discouraged. The validation set is formed of data which is weighted or selected based on task-specific quality, either using on human labels or controller predictions. This ensures that selection based on task-specific quality, in the train set, is encouraged as opposed to selecting all samples to improve generalisability.
It should be noted that other strategies may be used to compute $\tilde{R}_t$, however, in this work we only evaluate the three outlined above. 

To form the final reward $R_t$, we clip the performance measure $\tilde{R}_t$ using a clipping quantity $\bar{R}_t$. The final reward thus takes the form:
\begin{align}
    R_t=\tilde{R}_t-\bar{R}_t
\end{align}

Similar to $\tilde{R}_t$, several different formulations may be used for $\bar{R}_t$. In this work we use a moving average $\bar{R}_t={\alpha}_R\bar{R}_{t-1}+(1-{\alpha}_R)\tilde{R}_t$, where ${\alpha}_R$ is a hyper-parameter set to 0.9. It should be noted that, although optional, moving average-based clipping serves to promote continual improvement. A random selection baseline model or a non-selective baseline model may also be used as $\bar{R}_t$, however, these are not investigated in this work.

The RL-based meta-learning framework to learn IQA is summarised in Fig. \ref{fig:methods_fig} and is outlined in Algo. \ref{algo:iqa}.

\subsection{Learning task-agnostic image quality assessment}\label{sec:task_agnos_iqa}

As outlined in Sect. \ref{sec:intro}, under certain circumstances, it may be useful to learn task-agnostic IQA. For example when imaging protocols may need to be fine-tuned in order to remove noise and artefacts or when a particular target task may not be known. It is possible to learn such a task-agnostic IQA in the framework presented in Sect. \ref{sec:iqa_rl} using auto-encoding as the target task. It should be noted that while auto-encoding or self-reconstruction is a target task in this RL framework, it is not a clinical task and is only used for the purpose of learning a task-agnostic IQA.

For the auto-encoding target task, the label distribution is set as the image distribution $\mathcal{Y}=\mathcal{X}$. With the $L_h$ based on image reconstruction error, such as mean squared error $1/n \sum_{i=1}^{n} (y_t^i - \hat{y}_t^i)^2$ where $\hat{y}$ is the task predictor-predicted label, in the presented framework. Features not common across the entire distribution $\mathcal{X}$ such as random noise and randomly placed artefacts may be difficult to reconstruct \citep{autoenc_denoise_1, autoenc_denoise_2, autoenc_denoise_med}. The intuition is that due to a higher reconstruction error for such samples, these samples are to be valued lower by the controller. In addition to detecting samples with random noise and artefacts, this scheme may also be used for unsupervised anomaly detection although this is not discussed further in this work.

\subsection{Shaping task-specific rewards to weight task-agnostic measure}\label{sec:shaping}

Let $h_a(\cdot; \theta_a)$ be a pre-trained task-agnostic quality-predicting controller, which has been trained as described in Sect. \ref{sec:task_agnos_iqa}. This fixed pre-trained task-agnostic quality-predicting controller is used to shape the task-specific rewards in order to learn a weighted or overlapping measure of image quality. A new controller can then be trained with a shaped reward signal where the weighting between the task-agnostic quality and the task-specific quality can be manually adjusted. We use the term `shaped' for this new reward since, as opposed to the reward formulation in Sect. \ref{sec:iqa_rl} with a single reward per-batch, the shaped formulation has per-sample rewards as outlined in Eq. \ref{eq:shaping}. A controller trained using the non-shaped reward may not be able to distinguish between samples that negatively impact task performance due to general quality defects, and samples that negatively impact performance due to clinical difficulty. The learned ability, using the shaped reward signal, to identify only samples that negatively impact task performance due to imaging artefacts or general quality defects may be clinically useful. This ability to distinguish such cases can help to identify samples with general quality defects that may need re-acquisition. As opposed to simply computing reward using $R_t = \tilde{R}_t - \bar{R}_t$, the shaped reward at time-step $t$ can be computed as follows:


\begin{align}\label{eq:shaping}
    R_t = \{\phi(\tilde{R}_t-\bar{R}_t) + (1-\phi)(h_a(x_t^i ;\theta_a))\}_{i=1}^P
\end{align}

Here $P=B+M$ is a set of samples formed by the the concatenation of the mini-batch of train samples (with $B$ samples) and the $M$ validation set samples. The per-sample reward for these $P$ samples is computed using Eq. \ref{eq:shaping}. The clipping, in this shaped reward, is only applied to $\tilde{R}_t$ (the task performance measure) and not to the entire reward since the task-agnostic IQA controller $h_a$ is fixed and its predictions are deterministic during training with the shaped reward. This weighted sum of the clipped task performance measure, for the task in question, and the task-agnostic IQA, allows for manual adjustment of the relative importance of the task-agnostic quality and task-specific quality. It should be noted that instead of using $h_a$ in this shaping strategy it is also possible to use human labels of IQA or a different task-specific quality-predicting controller which is specific to a different target task. After training with the shaped reward, the trained controller may be denoted as $h(\cdot; \theta)$. When $\phi=1$, the shaped reward simplifies to the non-shaped reward $R_t = \tilde{R}_t - \bar{R}_t$, which was introduced in Sect. \ref{sec:iqa_rl}, thus a fully task-specific IQA is learnt. When $\phi=0$, the shaped reward simplifies to $R_t = \{h_a(x_t^i ;\theta_a)\}_{i=1}^B$ which means that the trained controller will be approximately equal to the pre-trained task-agnostic quality-predicting controller $h(\cdot; \theta) \approx h_a(\cdot; \theta_a)$. In this work, for notational convenience, wherever we use $\phi=0$, we report results for $h_a(\cdot; \theta_a)$ directly and do not train $h(\cdot; \theta)$.

\begin{algorithm}[!ht]
    \SetAlgoLined
    \KwData{Training dataset $\mathcal{D}_\text{train}$ and validation dataset $\mathcal{D}_\text{val}$.}
    \KwResult{Task predictor $f(\cdot;w)$ and controller $h(\cdot;\theta)$.}
    \BlankLine
    \While{not converged}{
        \For{$k\leftarrow 1$ \KwTo $K$}{
            \For{$t\leftarrow 1$ \KwTo $T$}{
                Sample a mini-batch $\mathcal{B}_t=\{(x_t^i,y_t^i)\}_{i=1}^{B}$ from $\mathcal{D}_\text{train}$\;
                Compute selection probabilities $\{h(x_t^i;\theta_t)\}_{i=1}^B$\;
                Sample actions $a_t=\{a_t^i\}_{i=1}^B$ w.r.t. $a_t^i\sim\text{Bernoulli}(h(x_t^i; \theta))$\;
                Selected samples $\mathcal{B}_{t,\text{selected}}$ from $\mathcal{B}_t$\;
                Update predictor $f(\cdot;w_t)$\ using $\mathcal{B}_{t,\text{selected}}$\;
                Compute reward $R_t$\;
            }
            Collect one episode $\{\mathcal{B}_t,a_t,R_t\}_{t=1}^T$\;
        }
        Update controller $h(\cdot;\theta)$ using reinforcement learning algorithm\;
    }
\caption{Image quality assessment}\label{algo:iqa}
\end{algorithm}

\section{Experiments}

\subsection{Ultrasound image quality for prostate detection and segmentation}

\subsubsection{Prostate classification and segmentation networks and the controller architecture}

For the prostate presence classification task, which is a binary classification of whether a 2D US slice contains the prostate gland or not, the Alex-Net \citep{alexnet_class} was used as the task predictor with a cross-entropy loss function and reward based on classification accuracy (Acc.), i.e. classification correction rate. The controller was trained using the deep deterministic policy gradient (DDPG) RL algorithm \citep{ddpg} where both the actor and critic used a 3-layer convolutional encoder for the image before passing to 3 fully connected layers.

For the prostate gland segmentation task, which is a segmentation of the gland on a 2D slice of US, a 2D U-Net \citep{unet_seg} was used as the task predictor with a pixel-wise cross-entropy loss and reward based on mean binary Dice score (Dice). The controller training and architecture details are the same as the prostate presence classification task.

Further implementation details can be found in the GitHub repository:~\url{https://github.com/s-sd/task-amenability/tree/v1}.

\subsubsection{Evaluating the task amenability agent}

The experiments for the two tasks of prostate presence classification and gland segmentation have been previously described in \citet{saeed_amenability} and the results are summarised in Sect. \ref{sec:res}. For these experiments, the task-specific IQA was investigated, i.e. the agent was trained with $\phi=1$. Evaluation of the proposed task amenability agent for both tasks, including the three proposed reward strategies, and comparisons to human labels of IQA were presented in our previous work \cite{saeed_amenability}.

Acc. and Dice were used as measures of performance for the classification and segmentation tasks, respectively. These serve as direct measures of performance for the task in question and as indirect measures of performance for the controller with respect to the learnt IQA. Standard deviation (St.D.) is reported as a measure of inter-patient variance and T-test results are reported at a significance level of $\alpha=0.05$, where comparisons are made. Details on controller selection are as described in Sect. \ref{sec:exp_mr_ts_agent}.

\subsubsection{Experimental data}
\begin{figure}[!ht]
\centering
\includegraphics[width=0.99\textwidth]{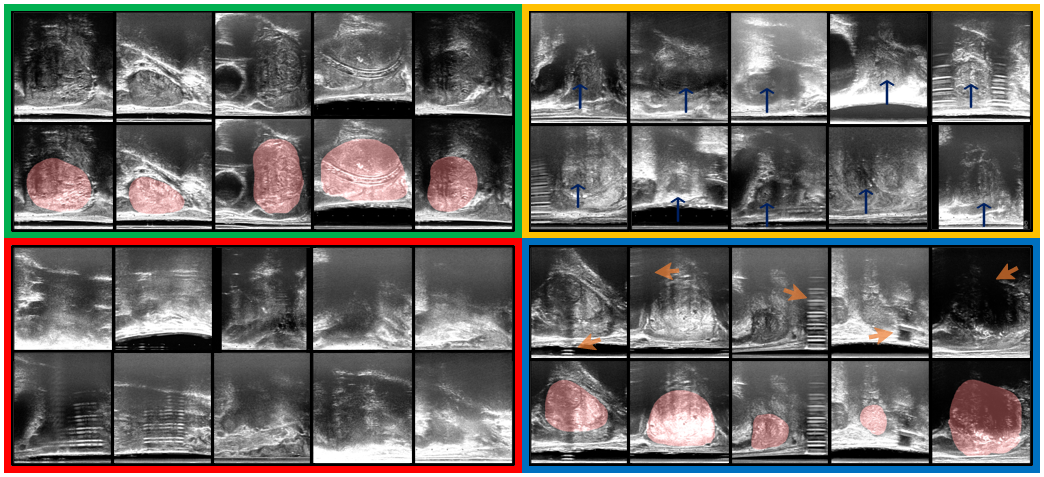}
\caption{Examples of ultrasound images used in this study. \textbf{Top-left} (green): task-amenable images that contain prostate gland (shaded in red); \textbf{Bottom-left} (red): images with poor task amenability where recognising the prostate (for classification) and delineating its boundary (for segmentation) is difficult; \textbf{Top-right} (yellow), images that likely contain prostate glands (blue arrows) but identifying the complete gland boundaries for segmentation is challenging; and \textbf{Bottom-right} (blue): images that contain visible noise and artefacts (orange arrows), but may be amenable to both classification and segmentation tasks.}
\label{fig:trus_samples}
\end{figure}

 The data was acquired as part of the SmartTarget:Biopsy and SmartTarget:Therapy clinical trials (NCT02290561, NCT02341677). Images were acquired form 259 patients who underwent ultrasound-guided prostate biopsy. In total 50-120 2D frames of TRUS were acquired for each patient using the side-firing transducer of a bi-plane transperineal ultrasound probe (C41L47RP, HI-VISION Preirus, Hitachi Medical Systems Europe). These 2D frames were acquired during manual positioning of a digital transperineal stepper (D\&K Technologies GmbH, Barum, Germany) for navigation or the rotation of the stepper with recorded relative angles for scanning the entire gland. The TRUS images were sampled at approximately every 4 degrees resulting in a total of 6712 images. These 2D images were then segmented by three trained biomedical engineering researchers. For the first task of prostate presence classification, a binary scalar indicating prostate presence was derived from the segmentation for all three observers and then a majority vote was conducted in order to form the final labels for used for this task. For the prostate gland segmentation, binary segmentation masks of the prostate gland were used and a majority vote at the pixel level was used to form the final labels used for this task. The images were split into three sets, train, validation and holdout, with 4689, 1023, and 1000 images from 178, 43, and 38 subjects, respectively. Samples from the TRUS data are shown in Fig. \ref{fig:trus_samples}.

\subsection{Multiparametric MR image quality for tumour segmentation}
\subsubsection{Prostate tumour segmentation network and the controller architecture}
A 3D U-Net \citep{3d_unet_seg} was used as the task predictor with a loss formed of equally weighted pixel-wise cross-entropy and Dice (i.e. $1-\text{Dice}$) losses and the reward was based on Dice. The reward formulation used for the computation of $\tilde{R}$ was the weighted formulation. A depth of 4 was used with the U-Net where 4 down-sampling and 4 up-sampling layers were used. Each of the down-sampling modules consisted of two convolutional layers with batch normalisation and ReLU activation, and a max-pooling operation. Analogously, de-convolutional layers are used instead in the up-sampling part. Skip connections were used to connect layers in encoding part to the corresponding layers in decoding part.

The input to the network was in the form of a 3-channel 3D image with the channels corresponding to T2-weighted, diffusion-weighted with high b-value and apparent diffusion coefficient MR images. The output is a one-channel probability map of the lesion for computing the loss, which can then be converted to a binary map using thresholding when computing testing results. Other network training details are described in Sect.~\ref{sec:exp_mr_ts_agent}.

For experiments with the prostate mpMR images, the deep deterministic policy gradient (DDPG) algorithm \citep{ddpg} was used as the RL algorithm for training with the only difference compared to experiments using the TRUS data being that the 3-layer convolutional encoders for the actor and critic networks used 3D convolutions rather than 2D. This agent may be considered as having $\phi=1$.

More implementation details can be found in the GitHub repository:~\url{https://github.com/s-sd/task-amenability/tree/v1}.

\subsubsection{Evaluating the task amenability agent}\label{sec:exp_mr_ts_agent}

Dice was used a direct performance measure to evaluate the segmentation task and as an indirect measure of performance for the controller with respect to learnt IQA. Standard deviation (St.D.) is reported as a measure of inter-patient variance. Wherever comparisons are made, T-test results with a significance level of 0.05 are reported. Where controller selection is applicable on the holdout set, samples are ordered according to their controller predicted values and the lowest $k\times100\%$ samples are removed, where $k$ is referred to as the holdout set rejection ratio. This rejection ratio is specified where appropriate. The results are presented in Sect. \ref{sec:res}.

\subsubsection{Evaluating the task-agnostic IQA agent}
With the prostate mpMR images, we compare the proposed task-agnostic IQA strategy, presented in Sect. \ref{sec:task_agnos_iqa}, with task-specific IQA learnt using the framework outlined in Sect. \ref{sec:iqa_rl}. To train the task-agnostic IQA agent, the actor and critic networks remain the same as in Sect. \ref{sec:exp_mr_ts_agent}. The task predictor, is a fully convolutional auto-encoder with 4 down-sampling and 4 up-sampling layers, with convolutions being in 3D. The loss was based on mean squared error (ie. $1/n \sum_{i=1}^{n} (y_t^i - \hat{y}_t^i)^2$) and the reward was based on mean absolute error (ie. $1/n \sum_{i=1}^{n} |y_t^i - \hat{y}_t^i|$) which serves as a measure of image reconstruction performance. This agent may be considered as having $\phi=0$.

\subsubsection{Evaluating the reward-shaping IQA networks}
We also evaluate the effect of different values of the weighting between the task-agnostic and task-specific IQA, $\phi$, when using the reward shaping strategy presented in Sect. \ref{sec:shaping}. Moreover, samples are presented to further qualitatively evaluate the learnt IQA. These results are presented in Sect. \ref{sec:res}. 

\subsubsection{Experimental data}
There were 878 sets of mpMR image data acquired from 850 prostate cancer patients. Patients data were acquired as part several clinical trials carried out at University College London Hospitals, with a mixture of biopsy and therapy patient cohorts, including SmartTarget \citep{hamid_smarttarget}, PICTURE \citep{simmons_picture}, ProRAFT \citep{orczyk_proraft}, Index \citep{dickinson_index}, PROMIS \citep{bosaily_promis} and PROGENY \citep{linch_progeny}. All trial patients gave written consents and the ethics was approved as part of the respective trial protocols. Radiologist contours were done manually and were included for all lesions with a likert score equal or greater than three as the radiological ground-truth in this study. All the data volumes were resampled to $0.5\times{0.5}\times{0.5} mm^3$  in 3D space. A region of interest (ROI) with a volume size of $192\times{192}\times{96}$ voxels centring at the prostate was cropped according to the prostate segmentation mask. All the volumes were then normalised to an intensity range of [0,1]. The 3D image sequences used in our study include those with T2-weighted, diffusion-weighted with b values of 1000 or 2000, and apparent diffusion coefficient MR, all available. These were then split into three sets, the train, validation and holdout sets with 445, 64 and 128 images in each set, respectively, where each image corresponds to a separate patient.

\section{Results}\label{sec:res}

\subsection{Evaluation of the task amenability agent}

\subsubsection{Prostate presence classification and gland segmentation from trans-rectal ultrasound data}

\paragraph{Evaluating the three reward strategies}
The target task performance for the prostate presence classification and gland segmentation tasks, on controller selected holdout set samples was investigated in \citet{saeed_amenability}. The system took approximately 12h to train on a single Nvidia Quadro P5000 GPU. For both tasks, for all three tested reward strategies, significantly higher performance was observed compared to the non-selective baseline (p-value$<$0.001 for all). Comparing the weighted reward formulation to the fixed clean validation set reward formulation, for the classification and segmentation tasks, no statistical significance was found (p-value$=$0.06 and 0.49 respectively). Contrastingly, comparing the selective reward formulation with the fixed clean reward formulation, statistical significance was observed, with the selective formulation showing inferior performance for both tasks (p-value$<$0.001 for all). The plots of task performance against rejection ratio are presented in Fig. \ref{fig:reject_IQA}, for both tasks. In addition to what was reported in \citet{saeed_amenability}, we include a non-selective baseline in these plots for comparison. The peak classification Acc. are 0.935, 0.932 and 0.913 at 5\%, 10\% and 5\% rejection ratios, for the fixed-, weighted- and selective reward formulations, respectively, while the peak segmentation Dice are 0.891, 0.893 and 0.866 at 20\%, 15\% and 20\% rejection ratios, respectively.

\paragraph{Investigating the impact of the validation set rejection ratio hyper-parameter for the selective validation set strategy}
In a first ablation study, we investigate the impact of the validation set rejection ratio $s^{rej}$, when using the selective validation set reward formulation with a holdout set rejection ratio of $0.15$, on the learnt IQA using the prostate gland segmentation task. Increasing $s^{rej}$ from $0.00$ to $0.30$ in increments of $0.05$, we observed a statistically significant improvement in performance for each step increase (p-value$<$0.01 for all) up to $s^{rej}=0.20$. Overall, performance (Dice) was improved from $0.827\pm0.011$ at $s^{rej}=0.00$ to $0.882\pm0.017$ at $s^{rej}=0.20$. Further increasing to a value of $s^{rej}=0.25$ and comparing with preceding value of $s^{rej}=0.20$ led to no significance being found (p-value$=$0.37). Another step increase, $s^{rej}=0.30$, led to lower performance compared to preceding value of $s^{rej}=0.25$, with statistical significance (p-value$<$0.01). 

\paragraph{Investigating the impact of changing RL algorithms}
In a second ablation study, we investigate the impact of the RL algorithm on the prostate presence classification task. For the weighted reward formulation with a holdout set rejection ratio of 0.05 we observed task performance (Acc.) of $0.929\pm0.013$ using the proximal policy optimisation (PPO) RL algorithm \cite{schulman_ppo} and no significance was found when comparing this to the DDPG algorithm which had a performance of $0.926\pm0.012$ (p-value=0.09). 

\paragraph{Comparing controller-predicted IQA with human IQA}
In addition to the plots of task performance against holdout set rejection ratio presented in Fig. \ref{fig:reject_IQA}, we also present contingency tables comparing controller-learnt task amenability with human labels of IQA in Fig. \ref{fig:cm}, for the prostate presence classification and gland segmentation tasks. As reported in \citet{saeed_amenability}, we summarise these results here since they offer interesting insights into the learnt IQA compared to human-judged IQA. For the purpose of comparison, 0.05 and 0.15 of the holdout set are considered to have low controller-predicted values, for the classification and segmentation tasks, respectively. For these comparisons, Cohen's kappa values of 0.75, 0.51 and 0.30 were found in the classification task, for the fixed clean, weighted- and selective reward formulations, respectively, and with respective kappa values of 0.63, 0.48 and 0.37 obtained in the segmentation task.

\begin{table*}[!ht]
\centering
\caption{Results on the controller-selected holdout set (holdout set rejection ratio of 0.10 used).}
\begin{tabular}{|c|c|c|}
\hline
Task & Reward computation strategy & Mean $\pm$ St.D.\\
\hline
\multirow{8}{6em}{Lesion segmentation (Dice)} & Non-selective baseline & 0.354 $\pm$ 0.016\\
\cline{2-3}
& $\tilde{R}_{\text{w},t}$, weighted validation set, shaped reward ($\phi=0.00$) & 0.367 $\pm$ 0.017\\
\cline{2-3}
& $\tilde{R}_{\text{w},t}$, weighted validation set, shaped reward ($\phi=0.80$) & 0.375 $\pm$ 0.016\\
\cline{2-3}
& $\tilde{R}_{\text{w},t}$, weighted validation set, shaped reward ($\phi=0.85$) & 0.380 $\pm$ 0.021\\
\cline{2-3}
& $\tilde{R}_{\text{w},t}$, weighted validation set, shaped reward ($\phi=0.90$) & 0.388 $\pm$ 0.022\\
\cline{2-3}
& $\tilde{R}_{\text{w},t}$, weighted validation set, shaped reward ($\phi=0.95$) & 0.366 $\pm$ 0.018\\
\cline{2-3}
& $\tilde{R}_{\text{w},t}$, weighted validation set, shaped reward ($\phi=0.99$) & 0.405 $\pm$ 0.019\\
\cline{2-3}
& $\tilde{R}_{\text{w},t}$, weighted validation set, shaped reward ($\phi=1.00$) & 0.415 $\pm$ 0.020\\
\cline{2-3}
\hline
\end{tabular}
\label{tab:res_IQA}
\end{table*}

\begin{table}[!ht]
\centering
\caption{Dice results for varying $\phi$ values against varying values of $k$.}
\label{tab:res_shaping}
\begin{tabular}{|c|c|c|c|c|c|}
\hline
 \multicolumn{2}{|c|}{} & \multicolumn{4}{c|}{$\phi$} \\
\cline{3-6}
 \multicolumn{2}{|c|}{} & 0.00 & 0.85 & 0.95 & 1.00 \\
\hline
\multirow{6}{*}{$k$} & 0.00 & 0.354 $\pm$ 0.019 & 0.368 $\pm$ 0.020 & 0.377 $\pm$ 0.023 & 0.375 $\pm$ 0.017 \\
\cline{2-6}
 & 0.05 & 0.358 $\pm$ 0.018 & 0.381 $\pm$ 0.016 & 0.383 $\pm$ 0.019 & 0.397 $\pm$ 0.019 \\
\cline{2-6}
 & 0.10 & 0.367 $\pm$ 0.017 & 0.380 $\pm$ 0.021 & 0.396 $\pm$ 0.018 & 0.415 $\pm$ 0.020 \\
\cline{2-6}
 & 0.15 & 0.370 $\pm$ 0.019 & 0.383 $\pm$ 0.018 & 0.398 $\pm$ 0.016 & 0.408 $\pm$ 0.019 \\
\cline{2-6}
 & 0.20 & 0.368 $\pm$ 0.021 & 0.376 $\pm$ 0.016 & 0.394 $\pm$ 0.020 & 0.401 $\pm$ 0.015 \\
\cline{2-6}
 & 0.25 & 0.366 $\pm$ 0.017 & 0.375 $\pm$ 0.019 & 0.392 $\pm$ 0.017 & 0.404 $\pm$ 0.018\\
\hline
\end{tabular}
\end{table}

\subsubsection{Prostate lesion segmentation from mpMR images}

\begin{figure}[!ht]
\centering
\includegraphics[width=0.95\textwidth]{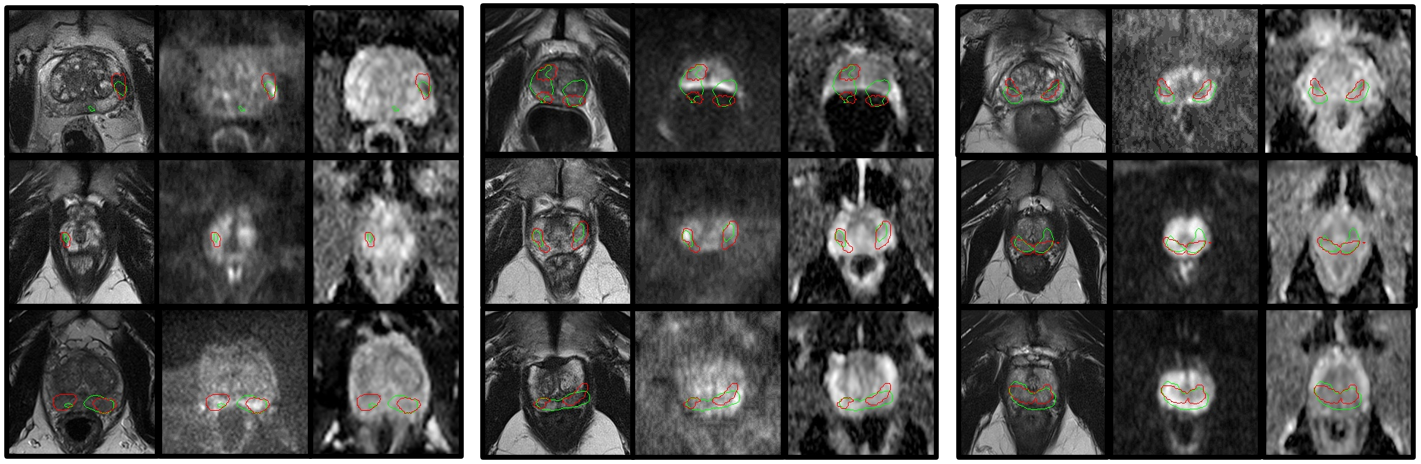}
\caption{Samples for the prostate lesion segmentation task with the three sub-columns in each column of samples being the three channels T2-weighted, diffusion-weighted, and apparent diffusion coefficient MR. Red is the ground truth and green is the predictor-predicted segmentation.}
\label{fig:lesion_seg_task_samples}
\end{figure}

\begin{figure}[!ht]
  \centering
\subfloat[Contingency tables comparing subjective labels (high or low task-specific quality labelled by an expert) to controller predictions (high or low task-specific quality determined by the controller using holdout set rejection ratios of 0.05 and 0.15 for the classification and segmentation, respectively) for the different reward formulations, from left to right: 1) fixed clean; 2) weighted; and 3) selective, validation sets. For comparison with controller-predicted ``ground-truth'' quality, number of true positive (TP), true negative (TN), false positive (FP) and false negative (FN) is summarised in the tables. For example, FN is the number of samples predicted as having high controller-predicted quality but low human observer-predicted quality.]{\includegraphics[width=0.65\textwidth]{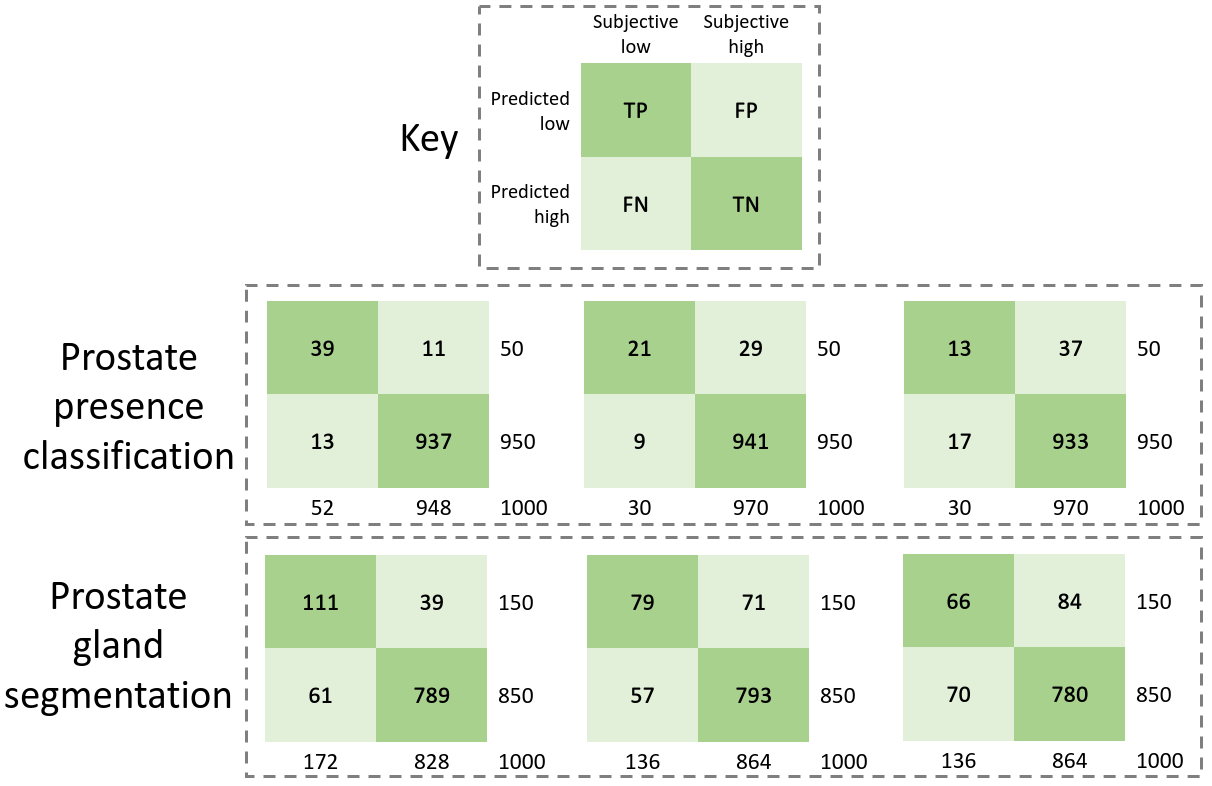}\label{fig:cm}}
\hfill
\subfloat[Plots of the task performance (in respective Acc. and Dice metrics) against the holdout set rejection ratio for the two tasks of prostate presence classification and gland segmentation.]{\includegraphics[width=0.31\textwidth]{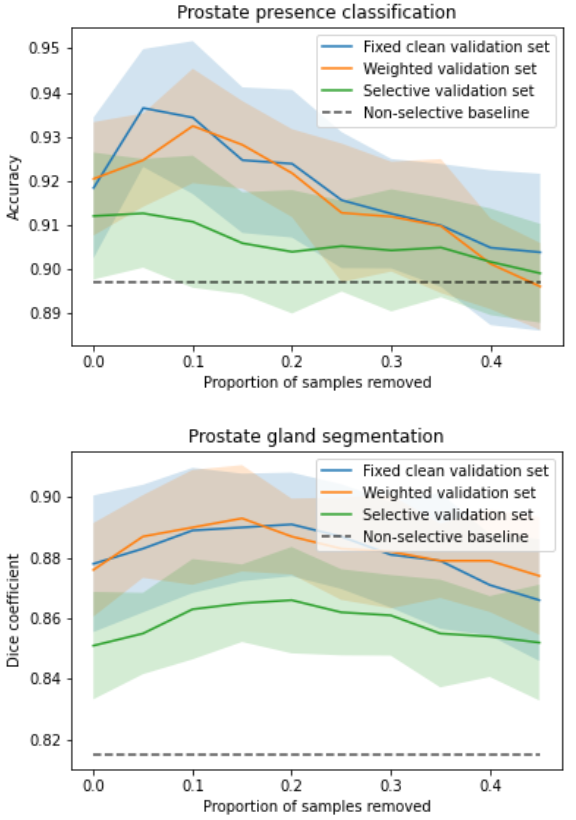}\label{fig:reject_IQA}}
    \caption{Summarised results for the prostate presence classification and prostate gland segmentation tasks for the TRUS dataset.}\label{fig:trus_res}
\end{figure}

\paragraph{Evaluating a fully task-specific reward}
The task performance results for the prostate lesion segmentation task, for controller-selected holdout samples, are presented in Table \ref{tab:res_IQA}. The approximate training time for the IQA system was 24h on a single Nvidia Tesla V100 GPU. The results for the task-amenability agent, where a fully task-specific IQA is learnt, are those with $phi=1$. For this agent, we observed higher performance for the controller-selected holdout set compared with the non-selective baseline (p$<$0.01). The plot of performance in terms of Dice against holdout set rejection ratio for this agent, presented in Fig. \ref{fig:lesion_seg_plot}, shows an initial rise followed by a plateau, as opposed to a small decrease after the initial rise which was observed for the prostate presence classification and gland segmentation tasks using the TRUS data. Samples of task predictor-predicted labels for the lesion segmentation task along with ground truth values are presented in Fig. \ref{fig:lesion_seg_task_samples}.



\subsection{Evaluation of the task-agnostic IQA agent}

\subsubsection{Prostate lesion segmentation from mpMR images}

The results for the task-agnostic IQA agent, for the lesion segmentation task using mpMR images, are summarised in Table \ref{tab:res_IQA} and Fig. \ref{fig:lesion_seg_iqa}. The task-agnostic IQA agent refers to the agent trained with $\phi=0$. Controller selection of holdout samples for this agent shows a small yet statistically significant improvement in performance compared to the non-selective baseline (p$<$0.01). A contingency table comparing task-agnostic IQA ($\phi=0.00$) with task-specific IQA ($\phi=1.00$) shows the level of disagreement between the two, with a Cohen's kappa value less than 0.001. Samples for controller predictions are presented in Fig. \ref{fig:lesion_seg_samples}.

\subsection{Evaluation of the reward shaping strategies}

\subsubsection{Prostate lesion segmentation from mpMR images}

\begin{figure}[!ht]
  \centering
  \subfloat[Contingency table comparing task-agnostic and task-specific IQA with samples determined high or low quality by each controller, with a holdout set rejection ratio of 0.10.]{\includegraphics[width=0.45\textwidth]{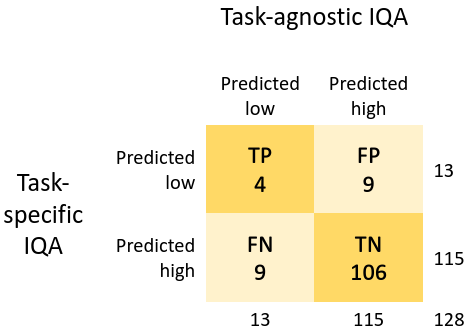}\label{fig:lesion_seg_conf_mat}}
  \hfill
  \subfloat[Plot of task performance (in Dice) against holdout set rejection ratio.]{\includegraphics[width=0.45\textwidth]{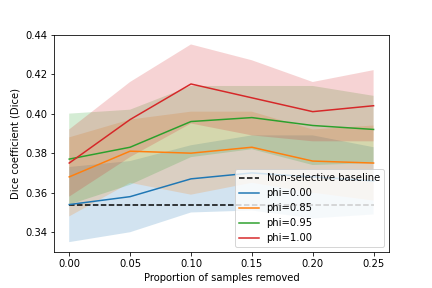}\label{fig:lesion_seg_plot}}
  \caption{Summarised results for the prostate lesion segmentation task.}
  \label{fig:lesion_seg_iqa}
\end{figure}

The results for the shaped reward formulations, for the lesion segmentation task using mpMR images, are summarised in Table \ref{tab:res_shaping} and Fig. \ref{fig:lesion_seg_plot}. Higher performance was observed for the controller-selected holdout set compared with the non-selective baseline, for all tested values of $\phi$, where the differences were statistically significant (p$<$0.01 for all). Moreover, the shaped reward formulation with $\phi\geq0.85$ showed improved performance compared to the shaped reward with $\phi=0.00$ (task-agnostic IQA agent), with statistical significance (p$<$0.01 for all). Interestingly, comparing the the formulation with $\phi=0.00$ to the formulation with $\phi=0.80$, statistical significance was not found (p-value$=$0.051). 

To further qualitatively assess the proposed IQA agent in identifying the general task-agnostic quality issues, 20 example cases were shown blindly to an experienced radiologist. Among the 8 cases reported as high task-agnostic quality by the IQA agent, the radiologist agreed with 7, with one having ``minor artefact". Among the other 12 cases deemed low task-agnostic quality with varying task-specific quality by the IQA agent, the radiologist agreed with 3 having ``significant quality issues that may affect diagnosis" and 6 having ``minor quality issues that are unlikely to affect the diagnostic task", and disagreed with the other 3 having ``little quality issues".


\begin{figure}[!ht]
    \centering
    \includegraphics[width=0.99\textwidth]{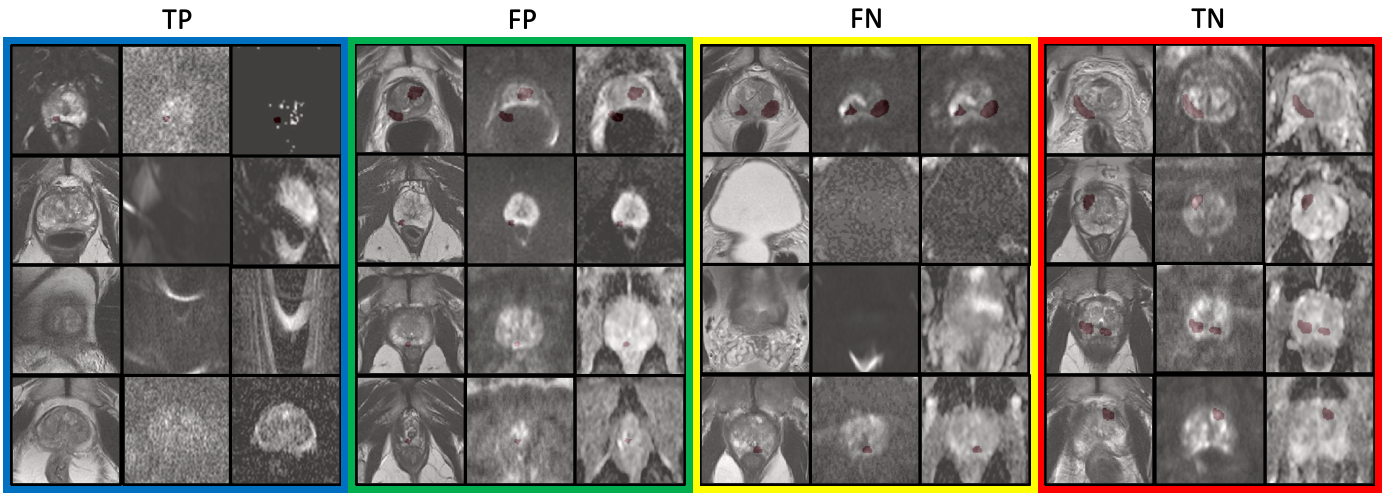}
\caption{Samples of mpMR images from the lesion segmentation task (see Fig. \ref{fig:lesion_seg_conf_mat} for definitions of TP, FN, FP and TN). In each coloured box, from left to right, the three channels of the same slice are shown (channels are T2-weighted MR, diffusion-weighted MR, and apparent diffusion coefficient MR). \textbf{TP} (Blue): Four slices form four different 3D volumes showing visible susceptibility and grain artefacts \citep{iqa_prostate_mr} (valued low by both the task-specific IQA and the task-agnostic IQA); \textbf{FP} (green): Four slices form four different 3D volumes showing a misaligned sample (third slice) and some samples with a very small lesion size with low tissue density contrast between lesion and surroundings (valued low by the task-specific IQA and high by the task-agnostic IQA); \textbf{FN} (yellow): Two slices from the one 3D volume (first and second slice) and two slices from another 3D volume (third and fourth slices) where the region which contains the lesion has no apparent artefacts but distortion artefacts are present in other slices (valued high by the task-specific IQA and low by the task-agnostic IQA); \textbf{TN} (red): Samples with no apparent artefacts (valued high by both the task-specific IQA and the task-agnostic IQA). It is also interesting to note that for the shaped reward formulation with $\phi=0.90$, the controller predicted values for these samples can be ordered from low to high in the following order: TP, FP, FN, TN (with samples at the top, in the same coloured box, being lower valued compared to samples on the bottom, if from two different 3D volumes).}
    \label{fig:lesion_seg_samples}
\end{figure}

\section{Discussion and Conclusion}

Results from the experiments investigating the three proposed reward strategies, summarised in Fig. \ref{fig:trus_res}, show that the selective reward formulation achieved inferior performance compared to the weighted and fixed clean validation set reward strategy. When tuning the $s^{\textbf{rej}}$ parameter for this selective formulation, however, we see a performance increase. Thus, further tuning of this parameter may be required in order to achieve performance comparable to the other reward formulations. Moreover, the selective formulation also offers a mechanism to specify a desirable validation set rejection ratio which may be useful for applications where there is a significant class imbalance problem. Additionally, the selective and weighted reward formulations allow for task-specific IQA to be learnt without any human labels of IQA and the weighted reward formulation does so without any significant reduction in performance compared to the fixed clean reward formulation, which requires human labels of IQA. In the experiments with the TRUS data, we see a trend where after an initial rise in performance with increasing holdout set rejection ratio, the performance slightly drops, this may be a dataset-specific phenomenon, since such drop was not observed in the lesion segmentation task on mpMR images. In the lesion segmentation task, after an initial rise, the performance seems to plateau with increasing holdout set rejection ratio. Nevertheless, some explanations for the plateau or small decrease may include high variance in the predictions, limited possible performance improvement due to the use of overall quality-controlled data from clinical trials, and over-fitting of the controller. 

As summarised in Fig. \ref{fig:lesion_seg_conf_mat} and \ref{fig:lesion_seg_samples}, the disagreement between the learnt task-specific IQA, for the prostate lesion segmentation task, and task-agnostic IQA, for the mpMR images, shows that learning varying definitions of IQA is possible within the proposed framework. It is interesting that while the task performance for the task-agnostic IQA was not comparable with the task-specific IQA, it was still able to offer improved performance compared to a non-selective baseline. This is potentially because it may be more difficult to perform any task on images which have a large amounts of defects. 

Manual adjustment of the trade-off between the task-specific and task-agnostic IQA, using the proposed reward shaping strategy, allows for different IQA definitions to be learnt which may be useful under different clinical scenarios. For given a scenario, a learnt definition can then be used to obtain relevant IQA scores for new samples. As an example, to inform re-acquisition decisions, it may be useful to set a threshold, on the controller-predicted IQA scores, where the IQA controller was trained using a shaped reward, such that newly acquired images that impact task performance negatively due to artefacts may be identifiable as opposed to identifying all samples that negatively impact performance regardless of the cause (when using fully task-specific IQA). In Fig. \ref{fig:lesion_seg_samples}, the ranking of the samples based on IQA with $\phi=0.90$ shows that it may be possible to define such a threshold using a holdout set rejection ratio such that only samples that impact task performance due to image quality defects such as grain, susceptibility and misalignment may be flagged for re-acquisition. The shaped reward thus provides a means to identify samples with quality defects that impact task-performance. This is in contrast to fully task-specific IQA which identifies all samples that impact task performance, regardless of the cause, and also to fully task-agnostic IQA which identifies samples that have quality defects, regardless of impact on task performance. 

The classification of samples presented in Fig. 7 shows that the task-specific IQA and task-agnostic IQA learn to valuate samples differently. The low-quality samples flagged by the fully task-specific IQA, i.e. TP and FP, appear to either have imaging defects including distortion, grain or susceptibility, or appear to be clinically challenging e.g. samples with small lesions. It should be noted that, however, on its own the fully task-specific IQA cannot distinguish between samples with artefacts and those that are clinically challenging. The low-quality samples flagged by the fully task-agnostic IQA, i.e. TP and FN, appear to either contain artefacts within (TP) or outside (FN) regions of interest such that they do not appear to impact the target task. Contrastingly, when we observe the valuation based on $\phi=0.90$ we see that, in FP, samples with quality defects also impacting the target task (TP), e.g. problems within regions of interest, are valued lowest; and sequence-misaligned samples are valued lower compared to clinically challenging samples. These are examples indicating that setting a threshold on the controller with shaped reward is effective, such that samples with defects impacting the target task can be identified for re-acquisition.

While the overlapping measures of IQA do not achieve the best average performance, they provide a means to identify samples for which performance may be improved by re-acquisition or defect correction, such as artefact removal or de-noising. This is because re-acquisition of samples with defects that do not impact the task is not beneficial and potentially expensive (e.g. samples with artefacts outside regions of interest). On the other hand, re-acquisition of clinically challenging samples is futile since performance for these samples cannot be improved by re-acquisition (e.g. samples with small tumour size). By definition, the fully task-specific IQA achieves highest performance (assessing based on simple sample removal), since it removes both clinically challenging samples and samples which impact the task due to imaging defects, without any distinction between the two. In contrast, the overlapping measure with the shaped reward can identify samples that impact the target task due to potentially correctable imaging defects such that if these defects were to be corrected, a performance improvement may be seen for the particular samples. For re-acquisition decisions, the overlapping measure also provides a possibility to identify samples such that target task performance, for those samples, may be improved if they are re-acquired.

It is also interesting that the task-agnostic IQA was able to identify samples which have quality defects such as distortion but which do not impact the target task due to being out of plane of the tumour. While this information may not be directly useful for the lesion segmentation task, it provides insight into types of imaging parameters or protocols, that are less important to a specific diagnosis task, for a more efficient and streamlined clinical implementation in the future.

It is important to highlight that the overall ranking of task amenability on a set of images will be altered using the combinatory IQA considering both types of qualities, compared with that from a purely task-specific IQA. The potential alternative strategy would be using the task-agnostic IQA on the subset of images selected by the task-specific IQA, with respect to a pre-defined threshold on task-specific image quality, or vice versa for a different potential application. Adjusting the reward shaping hyper-parameter at training time, using the scheme proposed in this work, is capable of achieving equivalent selections without the need for per-sample adjustment of thresholds. Adjusting these thresholds may also be inefficient during training time when individual or small batch of images are assessed. 
The combinatory IQA can, therefore, be used to output controller scores in a single forward pass, for new samples, where a threshold can be specified at training to produce re-acquisition decisions. Using either measure separately or in two sequential stages requires adjustment of holdout set rejection ratio thresholds for both the task-agnostic and task-specific qualities, which in addition to potentially requiring per-sample adjustment of the thresholds, would also require two forward passes through two separate controllers.

In this work, in addition to summarising the framework to learn task-specific IQA, previously presented in \citet{saeed_amenability}, we have presented a mechanism which allows for task-agnostic IQA to be learnt without any human labels of quality. Moreover, the reward shaping mechanism is proposed with a manually adjustable trade-off between the task-specific and task-agnostic IQA which may be tuned for a wide range of potential applications. These extended methodologies were evaluated using a diagnostic target task of prostate lesion segmentation using mpMR images acquired from clinical prostate cancer patients.



\acks{
This work was supported by the International Alliance for Cancer Early Detection, an alliance between Cancer Research UK [C28070/A30912; C73666/A31378], Canary Center at Stanford University, the University of Cambridge, OHSU Knight Cancer Institute, University College London and the University of Manchester. This work is supported by the Wellcome/EPSRC Centre for Interventional and Surgical Sciences [203145Z/16/Z]. F. Giganti is a recipient of the 2020 Young Investigator Award funded by the Prostate Cancer Foundation / CRIS Cancer Foundation. Z.M.C. Baum is supported by the Natural Sciences and Engineering Research Council of Canada Postgraduate Scholarships-Doctoral Program, and the University College London Overseas and Graduate Research Scholarships. Research reported in this publication was supported by the National Cancer Institute of the National Institutes of Health under Award Number R37CA260346. The content is solely the responsibility of the authors and does not necessarily represent the official views of the National Institutes of Health.
}

%
\ethics{The work follows appropriate ethical standards in conducting research and writing the manuscript, following all applicable laws and regulations regarding treatment of animals or human subjects.}

\coi{We declare that we do not have conflicts of interest.}

\bibliography{sample}




\end{document}